\documentclass[twocolumn,floatfix, showpacs]{revtex4}

\usepackage[final]{epsfig}
\usepackage{amsmath}

\begin{document}
 
\title{On the sensitivity of the dijet asymmetry to the physics of jet quenching}
 
\author{Thorsten Renk}
\email{thorsten.i.renk@jyu.fi}
\affiliation{Department of Physics, P.O. Box 35, FI-40014 University of Jyv\"askyl\"a, Finland}
\affiliation{Helsinki Institute of Physics, P.O. Box 64, FI-00014 University of Helsinki, Finland}


\begin{abstract}
The appearance of monojets is among the most striking signature of jet quenching in the context of ultrarelativistic heavy-ion collisions. Experimentally, the disappearance of jets has been quantified by the ATLAS and CMS collaborations in terms of the dijet asymmetry observable $A_J$. While the experimental findings initially gave rise to claims that the measured $A_J$ would challenge the radiative energy loss paradigm, the results of a systematic investigation of $A_J$ in different models for the medium evolution and for the shower-medium interaction presented here suggest that the observed properties of $A_J$ arise fairly generically and independent of specific model assumptions for a large class of reasonable models. This would imply that rather than posing a challenge to any particular model, the observable prompts the question what model dynamics is not compatible with the data.
\end{abstract}
 
\maketitle

\section{Introduction}

Hard partonic reactions in perturbative Quantum Chromodynamics (pQCD) lead (to leading order in the strong coupling $\alpha_s$) to highly virtual back-to-back parton configurations which subsequently evolve as parton showers, hadronize and are then experimentally observed as back-to-back jet events. One of the most striking manifestations of the interaction of hard partons with a QCD medium as created in ultrarelativistic heavy-ion collisions (URHIC) is the appearance of monojets in events in which one of the outgoing parton showers has lost such a high amount of energy into the medium that it is no longer identified by a jet finding algorithm.

Such monojets have been observed in experiment indirectly via triggered two-particle correlations at RHIC energies e.g. by the STAR collaboration \cite{STAR-DzT}, but only recently directly on the level of reconstructed jets by the ATLAS and CMS collaborations at LHC energies \cite{ATLAS,CMS}.

The LHC experiments quantify the observation by focusing on the imbalance of the energy reconstructed for two jets in opposite hemispheres. Specifically, in the case of the ATLAS experiment, calorimeter towers are clustered within a jet cone radius of $R<0.4$, a trigger condition on the energy $E_{T1}$ of the most energetic jet of $E_{T1} > 100$ GeV is imposed and for valid trigger the energy of the leading jet in the opposite hemisphere $E_{T2}$ is measured with the additional condition $E_{T2} > 25$ GeV imposed. 

The asymmetry or dijet imbalance is then defined as

\begin{equation}
A_J = \frac{E_{T1} - E_{T2}}{E_{T1} + E_{T2}}
\end{equation}

where $A_J = 0$ indicates a perfect energy balance between the two jets whereas $A_J = 1$ indicates a true monojet event (note that the condition $E_{T2} > 25$ GeV prevents this situation in the ATLAS analysis).

A number of models which were able to account for the measured imbalance were subsequently proposed. Such models range from parametric estimates \cite{Collimation} via schematic modelling of energy loss and jet finding \cite{Dijets-Qin} to sophisticated NLO pQCD modelling of jet evolution in a medium, however without detailed medium or jet-finding modelling \cite{Dijets-Vitev} and complete Monte Carlo (MC) modelling of a jet embedded into a bulk medium with realistic jet finding using the MARTINI code \cite{Dijets-Martini}. The success of these models quickly established that the measured dijet asymmetry does not pose a fundamental challenge to the radiative energy loss picture.

However, at least an equally interesting question is what the information content of the $A_J$ measurement is, i.e. what model properties are constrained by the observable. The aim of this work is to address this question by computing $A_J$ under a number of different assumptions for both the parton-medium interaction and the evolution of the bulk medium.

\section{Methodology}

A rigorous comparison of a jet quenching theory with e.g. the ATLAS measurement would in principle require the following steps: 1) generation of a bulk event using for instance hydrodynamics with event by event fluctuating initial conditions 2) generation of a hard event inside the same bulk event (since there is an observable correlation between medium fluctuation hotspots and hard vertices \cite{Fluctuations}, the hard and soft physics can not be modelled independently in a rigorous treatment) and evolution of the perturbative shower inside the medium 3) computation of the back-reaction of the shower to the medium to account for correlated soft physics inside the jet area (for instance elastic recoil of medium constituents) 4) simulating the detector response to turn particle information into calorimeter tower information 5) clustering and analysis of the calorimeter-level event as in the experimental procedure.

The full set 1) - 5) corresponds to a numerically very demanding workflow, and consequently in the most ambitious theory comparison so far \cite{DijetsMartini}, only part of this has been included consistently using the bulk matter evolution code MUSIC without fluctuations and the in-medium shower evolution code MARTINI \cite{Dijets-Martini}. In the present work, we do not aim at a rigorous comparison with the data but rather at establishing the information content of an $A_J$ measurement, therefore the following simplifications are assumed:

\begin{itemize}
\item the event by event fluctuations of the medium are (for the purpose of embedding and propagating hard partons) replaced by an initial-state averaged hydrodynamical model
\item no back-reaction of the medium to energy-deposition by the shower is included, all energy transferred to the medium is assumed to be outside the jet definition, i.e. 'lost'
\item a simple set of cuts on hadron species and $P_T$ replaces the mapping of particles to calorimeter towers
\item instead of the anti $k_T$ algorithm clustering the whole event, a simple cone drawn around the primary parton direction with radius $R<0.4$ is used to define the jet, the (by assumption) smooth medium background can be subtracted by definition
\item medium (and calorimeter) fluctuations are then re-introduced  as uncorrelated Gaussian fluctuations of the medium background energy inside the cone area
\end{itemize}

While these assumptions have been made to mimick the behaviour of the published ATLAS baseline simulation \cite{ATLAS}, the reader should be cautious to read anything more than qualitative trends into a comparison with data. Note however that qualitatively, the simplified jet definition contains the same elements as the real calorimetric jet problem, e.g. hadron-species dependence of the calorimeter response or a momentum cut introduced by the deflection of low $P_T$ charged hadrons by the magnetic field.

Inside this framework, we test three different scenarios for the parton-medium interaction and three different hydrodynamical models for the medium evolution. Out of each model class, two of three model are known ruled out by some other set of data and are only used to test if $A_J$ is able to discriminate the elements in which these models fail.

Since the various test cases are easily distinguished by other observables (such as the single hadron suppression factor $R_{AA}$), it is important to set up the calculation in such a way that only the information provided by $A_J$ is probed. In particular, the dijet asymmetry is related to a conditional probability, i.e. given a trigger $E_{T1} > 100$ GeV, what is the probability to find an away side jet with $E_{T2}$? We're specifically interested in the information content of this conditional probability, not in the probability to find a trigger jet per event (which is related to the suppression factor $R_{AA}^{jets}$ of jets) in the first place.

The dependence on the reduction of the trigger rate is removed as follows: The parameters of the only realistic model combination (YaJEM-DE in LHC hydrodynamics) are tuned such that the single inclusive charged hadron suppression factor $R_{AA}$ at 60 GeV is described (this implies typically partonic kinematics above 100 GeV as relevant for the dijet observation). The parameters of all other model combinations are then adjusted to create the same suppression of the jet $> 100$ GeV trigger rate, i.e. all scenarios by definition show the same suppression of the single jet rate. This procedure is very similar to what has been used to assess the information content of the dihadron correlation suppression factor $I_{AA}$ in \cite{Dihadron1,Dihadron2}.

\section{The model}

The modelling thus consists of the MC simulation of a hard partonic back-to-back event, the choice of a medium evolution and embedding of the hard event into the medium, the propagation of the evolving parton showers through the medium and the clustering and binning according to the jet definition. 

\subsection{The perturbative hard process}

In LO factorized pQCD, the production of two hard partons $k,l$ 
is described by

\begin{equation}
\label{E-2Parton}
  \frac{d\sigma^{AB\rightarrow kl +X}}{dp_T^2 dy_1 dy_2} \negthickspace 
  = \sum_{ij} x_1 f_{i/A}(x_1, Q^2) x_2 f_{j/B} (x_2,Q^2) 
    \frac{d\hat{\sigma}^{ij\rightarrow kl}}{d\hat{t}}
\end{equation}

where $A$ and $B$ stand for the colliding objects (protons or nuclei) and 
$y_{1(2)}$ is the rapidity of parton $k(l)$. The distribution function of 
a parton type $i$ in $A$ at a momentum fraction $x_1$ and a factorization 
scale $Q \sim p_T$ is $f_{i/A}(x_1, Q^2)$. The distribution functions are 
different for free protons \cite{CTEQ1,CTEQ2} and nucleons in nuclei 
\cite{NPDF,EKS98,EPS09}. The fractional momenta of the colliding partons $i$, 
$j$ are given by $ x_{1,2} = \frac{p_T}{\sqrt{s}} \left(\exp[\pm y_1] 
+ \exp[\pm y_2] \right)$.
Expressions for the pQCD subprocesses $\frac{d\hat{\sigma}^{ij\rightarrow kl}}{d\hat{t}}(\hat{s}, 
\hat{t},\hat{u})$ as a function of the parton Mandelstam variables $\hat{s}, \hat{t}$ and $\hat{u}$ 
can be found e.g. in \cite{pQCD-Xsec}.

To account for various effects, including higher order pQCD radiation, transverse motion of partons in the nucleon (nuclear) wave function and effectively also the fact that hadronization is not a collinear process, the distribution is commonly folded with an intrinsic transverse momentum $k_T$ with a Gaussian distribution, thus creating a momentum imbalance between the two partons as ${\bf p_{T_1}} + {\bf p_{T_2}} = {\bf k_T}$. 

These expressions are evaluated in a MC framework described in \cite{Dihadron1,Dihadron2}, resulting in an approximately back-to-back pair of hard partons with known parton types and momenta.

\subsection{Embedding into hydrodynamics}

For the modelling of the medium created in URHIC at the LHC, we use a 2+1d ideal hydrodynamical \cite{LHChydro2d} framework which has successfully been tested against bulk spectra and already been used as a background to compute single hadron suppression at LHC energies \cite{LHCspectra}. We contrast this with two hydrodynamical evolutions for RHIC energies, a 3+1d code \cite{hydro3d} and a 2+1d code \cite{hydro2d} which do clearly \emph{not} reproduce bulk observables at LHC correctly.

These frameworks differ substantially in the geometry of the simulated medium, for a comparison between 2+1d frameworks at RHIC and LHC see e.g. \cite{LHChydro2d}, for a comparison between 2+1d and 3+1d at RHIC see \cite{JetHydSys} and are known to be discriminated by other high $P_T$ observables, for instance the reaction plane angle dependence of single hadron suppression $R_{AA}(\phi)$ \cite{JetHydSys} or by hard dihadron correlations \cite{Dihadron2}.

For the actual embedding of parton showers into the medium, we assume that the distribution of vertices in the medium follows binary collision scaling as appropriate for a LO pQCD calculation. Thus, the probability density to find a vertex in the transverse plane is
\begin{equation}
\label{E-Profile}
P(x_0,y_0) = \frac{T_{A}({\bf r_0 + b/2}) T_A(\bf r_0 - b/2)}{T_{AA}({\bf b})},
\end{equation}
where the thickness function is given in terms of Woods-Saxon distributions of the the nuclear density
$\rho_{A}({\bf r},z)$ as $T_{A}({\bf r})=\int dz \rho_{A}({\bf r},z)$ and $T_{AA}({\bf b})$ is the standard nuclear overlap function $T_{AA}({\bf b}) = \int d^2 {\bf s}\, T_A({\bf s}) T_A({\bf s}-{\bf b})$ for impact parameter ${\bf b}$. We place the parton pair generated in the previous step at a probabilistically sampled vertex $(x_0,y_0)$ sampled from this distribution with a random orientation $\phi$ with respect to the reaction plane.

When embedding parton showers into the medium, hydrodynamical parameters need to be related to transport coefficients which are probed by the shower. In the following, this is always done by assuming the relevant coefficients $\hat{q}$ (virtuality growth per unit pathlength) or  $D$ (mean energy loss per unit pathlength) are related to the energy density of the medium $\epsilon$ via 

\begin{equation}
\label{E-qhat}
\hat{q}[D](\zeta) = K[K_D] \cdot 2 \cdot \epsilon(\zeta)^{3/4} (\cosh \rho(\zeta) - \sinh \rho(\zeta) \cos\psi).
\end{equation}

Here $\zeta$ is the position coordinate of a parton in the medium, $\rho$ is the transferse flow rapidity and $\psi$ the angle between parton propagation and flow direction (the last factor contains a correction for relative motion between parton and fluid element, for instance it involves Lorentz contraction). The only parameters which are not fixed by the hydrodynamical medium are the adjustible constants $K$ (or $K_D$) which dial the overall strength of the medium modification. These are determined by a comparison to reference data or by the requirement of a constant jet rate suppression as described above.

\subsection{In-medium shower}

The in-medium shower for each parton given the path through the medium determined in the previous step is computed using the MC code YaJEM \cite{YaJEM1,YaJEM2,YaJEM-D}. YaJEM is a framework to compute medium-induced changes to the kinematics of the evolving parton shower where the medium is chacacterized in a phenomenological way by transport coefficients such as $\hat{q}$ or $D$ . The reliance on transport coefficients means that YaJEM does not resolve the medium microscopically. It is based on the PYSHOW algorithm \cite{PYSHOW} and has an option to simulate coherent medium-induced radiation as well as an incoherent drag force (as characteristic e.g. for elastic collisions with the medium constituents).
In the following, we use YaJEM to set up three scenarios:

The default scenario (YaJEM-DE) is characterized by medium-induced radiation with a $\sim 10$\% component of elastic energy loss \cite{YaJEM-DE}. The minimum scale $Q_F$ down to which the shower is evolved in the medium is set by formation time arguments as \cite{YaJEM-D}
\begin{equation}
\label{E-Q0}
Q_F = \sqrt{E/L}
\end{equation}
where $E$ is the parton energy and $L$ the in-medium length. This creates a length and energy dependence of the medium interaction which is indicated by a large set of high $P_T$ observables \cite{ElossPhysics}. Here, the small elastic component is required by the low momentum strength of dihadron correlations which is overestimated without this contribution \cite{YaJEM-DE}.

In an alternative scenario (YaJEM), we set the scale down to which the shower is evolved to a constant $Q_F = 0.7$ GeV. This changes pathlength and energy dependence and leads to disagreement with the high $P_T$ observables both at RHIC \cite{YaJEM-D} and  LHC \cite{LHCspectra}.

For a second alternative scenario (YaJEM-E), we assume that the medium effect is only manifest via incoherent drag and has no induced radiation component, i.e. $\hat{q} = 0$. This changes observables like subjet fraction or jet shapes to a behaviour opposite to an induced radiation scenario \cite{JetShapes}--- while induced radiation \emph{widens} a jet, a pure drag force scenario \emph{narrows} the jet shape (cf. RAD vs. DRAG in \cite{JetShapes}). The pathlength dependence of such a purely incoherent scenario is also incompatible with data \cite{ElasticPhenomenology}.

\subsection{Jet clustering and binning}

The YaJEM output in terms of hadron jets is then clustered to compute $P(E_{jet}|E,i,x,y,\phi)$, i.e. the probability distribution for a given parton $i$ with energy $E$ produced at a position $(x,y)$ in the transverse plane propagating at an angle $\phi$ to be registered as a jet of energy $E_{jet}$ after passage through the medium.

$E_{jet}$ is defined as the sum of the energy of hadrons fulfilling the following conditions: a) at an angle of $R=0.4$ or less with the jet axis b) the hadron species is one of $\pi^+, \pi^-, \pi^0, K^+, K^-, p, \overline{p}, \gamma$ c) $P_T > 1$ GeV (similar criteria have been suggested in the context of jet measurements by the STAR collaboration \cite{Joern}). To account for uncorrelated fluctuations of the medium background, a $\Delta E$ is sampled from a Gaussian distribution centered around zero with width 20 GeV and is added to the jet energies in the presence of a medium. In \cite{BgFluct}, it has been pointed out that uncorrelated background fluctuations have a potentially strong influence on the observed dijet asymmetry and hence need to be accounted for.

In this way, the near and away side jet energies $E_{T1}$ and $E_{T2}$ can be computed from the partonic event and compared with the experimental trigger conditions. Given a triggered event, $A_J$ is then be evaluated and the procedure is repeated with the generation of the next partonic event.

\section{Results}

Before looking at the outcome of the calculation, let us briefly review what we might reasonably expect to see by a comparison with dihadron correlations. Given the same suppression of the trigger rate, triggered dihadron correlations are sensitive to the pathlength dependence of energy loss \cite{ElasticPhenomenology} and clearly able to rule out incoherent scenarios with linear pathlength dependence. They also have tomographic capability, i.e. they are able to distinguish two different hydrodynamical background models \cite{Dihadron2}. At the heart of these is a phenomenon commonly called \emph{surface bias} --- the distribution of vertices leading to a triggered event is very different from the unbiased binary collision distribution Eq.~(\ref{E-Profile}) and does not peak in the medium center but towards the medium surface \cite{Dihadron1}. This in turn forces a longer than average path for the away side parton, and the precise balance between surface bias and pathlength dependence of energy loss is reflected in the strength of the dihadron correlation. 

In addition, the for low associate momentum bins, the dihadron correlation is able to trace the flow of energy to subleading particles and establish a lower bound on the rate with which energy is transferred into the bulk fluid medium \cite{YaJEM-DE}. In principle, all these phenomena can also be realized in dijet correlations.

\subsection{The baseline}

Even in the absence of a medium, there is some imbalance between the jet energy seen on the near and away side. The strength of the imbalance depends on the jet definition --- energy-momentum conservation ensures that the energy is balanced (up to intrinsic $k_T$) when summing over the complete flow of energy and momentum resulting from the original parton evolution, and hence any imbalance is caused by part of the flow of energy and momentum going outside of the jet definition.

\begin{figure}[htb]
\begin{center}
\epsfig{file=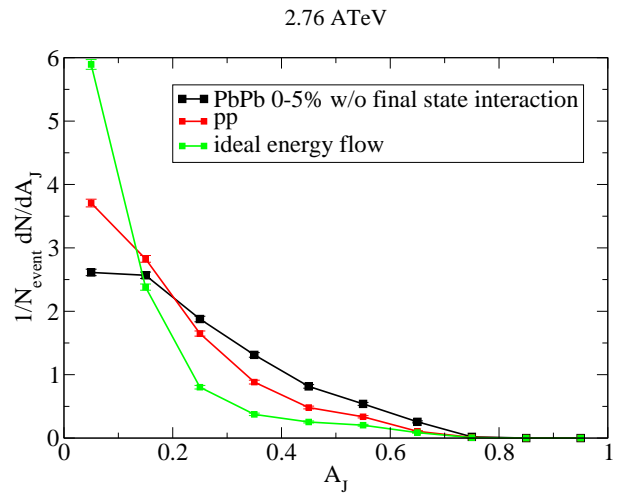, width=8cm}
\end{center}
\caption{\label{F-Base}(Color online) The dijet asymmetry $A_{jet}$ evaluated in the \emph{absence} of any parton medium interaction assuming that all energy within a cone of radius $R<0.4$ can be detected (ideal energy flow, green), applying the jet definition described in the text, but without bulk medium background fluctuations (pp), and in-medium baseline without final state interaction, but including uncorrelated Gaussian fluctuations of the medium background energy inside the jet area.}
\end{figure} 

As evident from Fig.~\ref{F-Base}, this is not an academic discussion --- the result with angular cut only and without any cuts on $P_T$ or hadron species is quite different from the result for the full in-medium baseline where the whole set of cuts and fluctuations has been applied. 

Note again that while the jet definition has been chosen to resemble the ATLAS baseline simulation results, there is no reason to think that it really corresponds to the ATLAS detector response and clustering procedure, however the implication is that the probability distribution for losing energy out of the jet definition is similar.

\subsection{Dependence on shower-medium interaction scenario}

\begin{figure}[htb]
\begin{center}
\epsfig{file=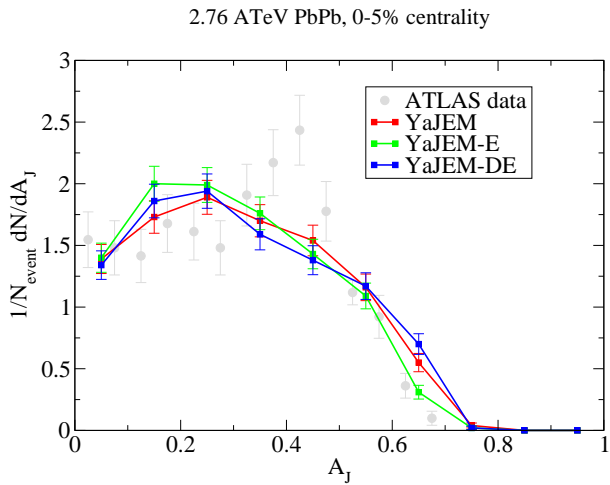, width=8cm}
\end{center}
\caption{\label{F-Model}(Color online) The dijet asymmetry $A_{jet}$ computed in the 2+1d hydrodynamical framework for LHC \cite{LHCspectra} for three different scenarios of the interaction of the evolving shower with the medium (see text). Shown is also ATLAS data \cite{ATLAS} for reference, note that however the jet definitions are not identical.}
\end{figure} 

The dependence of the distribution of $A_J$ on the assumed scenario for shower-medium interaction is shown in Fig.~\ref{F-Model}. Within statistical and systematic uncertainties, all three scenarios are indistinguishable and the width of the distibution qualitatively agrees with the ATLAS data. Given that both the pathlength dependence and the energy-momentum flow inside the modified jet is very different in all three scenarios, this is a somewhat remarkable outcome. The dijet asymmetry observable is neither strongly sensitive to the fact that YaJEM and YaJEM-E have a linear pathlength dependence which is incompatible with the angular dependence of single hadron suppression \cite{YaJEM-D} nor to the difference between widening and focusing of the jet shape \cite{JetShapes}. Only for large asymmetries $A_J > 0.6$ there is a hint of different model dynamics, indicating that a purely elastic energy loss scenario into the medium is somewhat less efficient in creating large asymmetries.

The suppression of the trigger jet rate as compared to the vacuum case as found in these scenarios corresponds to a suppression factor of $R_{AA}^{jets} = 0.48 \pm 0.02$ in the trigger momentum range (i.e. for jets of 100 to about 150 GeV), in line with ATLAS results for $R_{CP}$ \cite{ATLAS_talk}.

\subsection{Dependence on the hydrodynamical  evolution}

\begin{figure}[htb]
\begin{center}
\epsfig{file=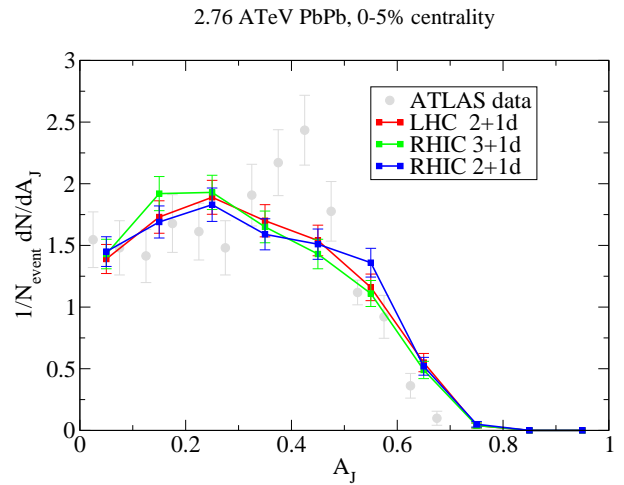, width=8cm}
\end{center}
\caption{\label{F-Hydro}(Color online) The dijet asymmetry $A_{jet}$ computed with YaJEM-DE in three different hydrodynamical evolution scenarios for the background. Shown is also ATLAS data \cite{ATLAS} for reference, note that however the jet definitions are not identical.}
\end{figure} 

In Fig.~\ref{F-Hydro} the dependence on the assumed hydrodynamial modelling is shown. Within statistical uncertainties, there is no such dependence visible despite the obvious differences in geometry, evolution time and freeze-out conditions when going from LHC to RHIC energies. Note however that by definition the differences in mean density between RHIC and LHC are not probed here, since in all cases the suppression of the trigger rate is required to be a constant. This implies that the quenching power scales ($K$ and $K_D$) are adjusted to a much higher value for the same value of the energy density in the RHIC hydrodynamical runs. If the same relation between medium density and transport coefficients would be assumed in all cases, the results would be very different, but note that such an overall normalization factor is much easier probed in single hadron suppression.

The obvious similarity of the outcome leaves only the conclusion that $A_J$ is not sensitive to the medium geometry, i.e. the observable has no tomographic capability.

\subsection{Dependence on other factors}

\begin{figure}[htb]
\begin{center}
\epsfig{file=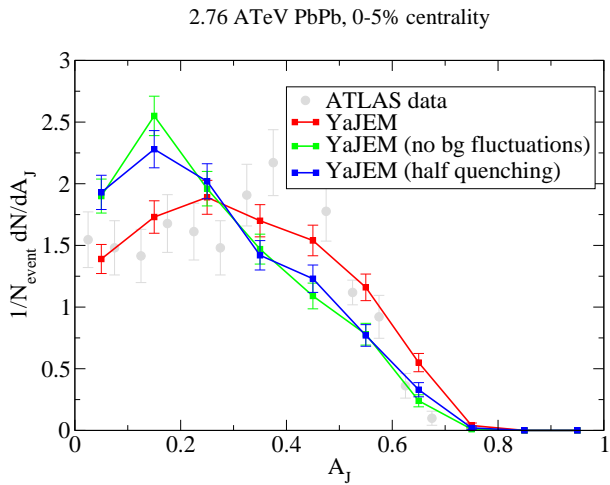, width=8cm}
\end{center}
\caption{\label{F-Fluct}(Color online) The dijet asymmetry $A_{jet}$ computed with YaJEM, assuming no fluctuations of the background or assuming only 50\% of the medium quenching power. Shown is also ATLAS data \cite{ATLAS} for reference, note that however the jet definitions are not identical.}
\end{figure} 

The dependence of $A_J$ on two other factors is illustrated by Fig.~\ref{F-Fluct}. If no fluctuations of the uncorrelated background medium inside the jet area are applied, then the shape of the $A_J$ distribution is found to be less broad. Similarly, if the medium transport coefficients are set to 50\% of the values used in the previous investigations, the resulting shape is quite different. However, note that both procedures alter the rate suppression of triggered events and are therefore not on par with the dependencies investigated in the previous sections.

These results demonstrate that the dijet asymmetry is sensitive to the overall strength of shower-medium interaction and that likewise that the proper treatment of the uncorrelated background fluctuations is important.

\subsection{Dependence on jet cone radius}

One of the experimental handles to control the amount of background fluctuations is the jet cone radius, and hence the systematics of the calculation with $R$ is an important crosscheck. Since the radius parameter controls the jet area and hence the total amount of background probed, the effect of uncorrelated background fluctuations should scale linearly with the radius parameter, i.e. clustering jets with $R=0.2$ should reduce background fluctuations by 50\% of the strength seen at $R= 0.4$.
In the following, we use this scaling assumption for the uncorrelated background.

However, while reducing the cone radius reduces background fluctuations, at the same time the probability for energy to fall outside the jet definition is increased. As a result, the fluctuations in clustered jet energy given a certain parton energy increase as compared to the case of $R=0.4$ (in principle even in the absence of a medium, but certainly in a more pronounced way in a medium). 

The net result of a variation in cone size depends on how well these two opposing effects cancel, since $A_J$ is in the end a measure of all jet energy fluctuations regardless of their origin. 

\begin{figure}[htb]
\begin{center}
\epsfig{file=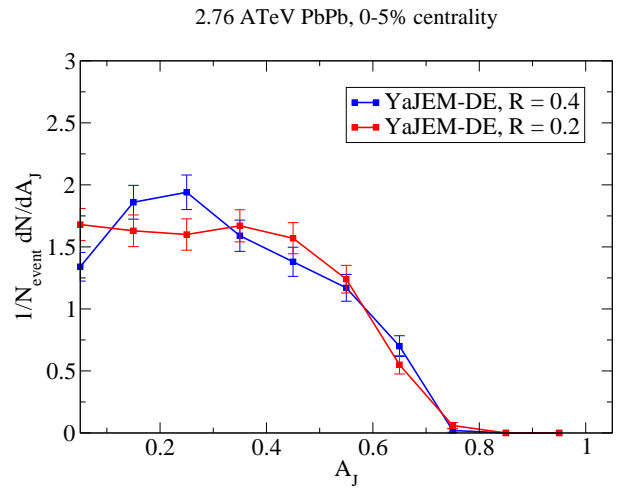, width=8cm}
\end{center}
\caption{\label{F-Radius}(Color online) The dijet asymmetry $A_{jet}$ computed with YaJEM-DE for two different choices of the jet cone radius, $R = 0.4$ (the default selection) and $R= 0.2$. }
\end{figure} 

The results using YaJEM-DE in the LHC 2+1d hydrodynamical evolution are shown in Fig.~\ref{F-Radius}. To first approximation the result indicates that the decrease of background fluctuations with $R$ is compensated by the increase in fluctuations of jet energy inside the cone. However, this cancellation isn't perfect and leads to small diffferences at low values of $A_J$.

For the same reason, also the suppression of the trigger rate is almost unchanged between $R = 0.4$ and $R=0.2$. For the smaller cone radius, the calculation yields $R_{AA}^{jets} = 0.46 \pm 0.02$ for jets in the trigger range, i.e. from 100 GeV to about 150 GeV, in line with the ATLAS results \cite{ATLAS_talk}.

Note that this cancellation of background against jet energy fluctuations inside the cone can not occur in the absence of background fluctuations. The fact that experimentally similar $R_{CP}$ and $A_J$ is observed for both $R=0.4$ and $R=0.2$ is a strong indication that both types of fluctuations are present, strengthening the message of \cite{BgFluct} that uncorrelated background fluctuations need to be properly understood.

\section{Discussion}

The apparent insensitivity of the dijet asymmetry observable $A_J$ to even gross changes in shower-medium interaction scenario or medium geometry is certainly at first glance more than somewhat surprising. However, upon closer inspection, several points can be identified which may help to explain the findings.

\subsection{The effect of clustering} 

Jet definitions as used in jet finding algorithms are designed to make meaningful comparison between pQCD calculations valid at large momentum scales $Q$ and observed hadron showers in terms of e.g. calorimetric measurements possible. For this purpose, jet definitions are created to systematically remove dependence to physics at low $Q$ where pQCD starts breaking down, i.e. by construction jet definitions are relatively blind to enhanced soft gluon emission or the dynamics of hadronization.

However, the medium modifications of a shower take place precisely at scales of the medium temperature, which in itself is not far from the QCD fundamental scale, $Q \sim T \sim \Lambda_{QCD}$. The implication is that vacuum jet finding is designed to minimize the sensitivity to physics at the medium scale, and as a result jets are fairly robust against final state interaction with the medium when compared to e.g. leading high $P_T$ hadrons.

\subsection{Lack of surface bias}

\begin{figure}[htb]
\begin{center}
\epsfig{file=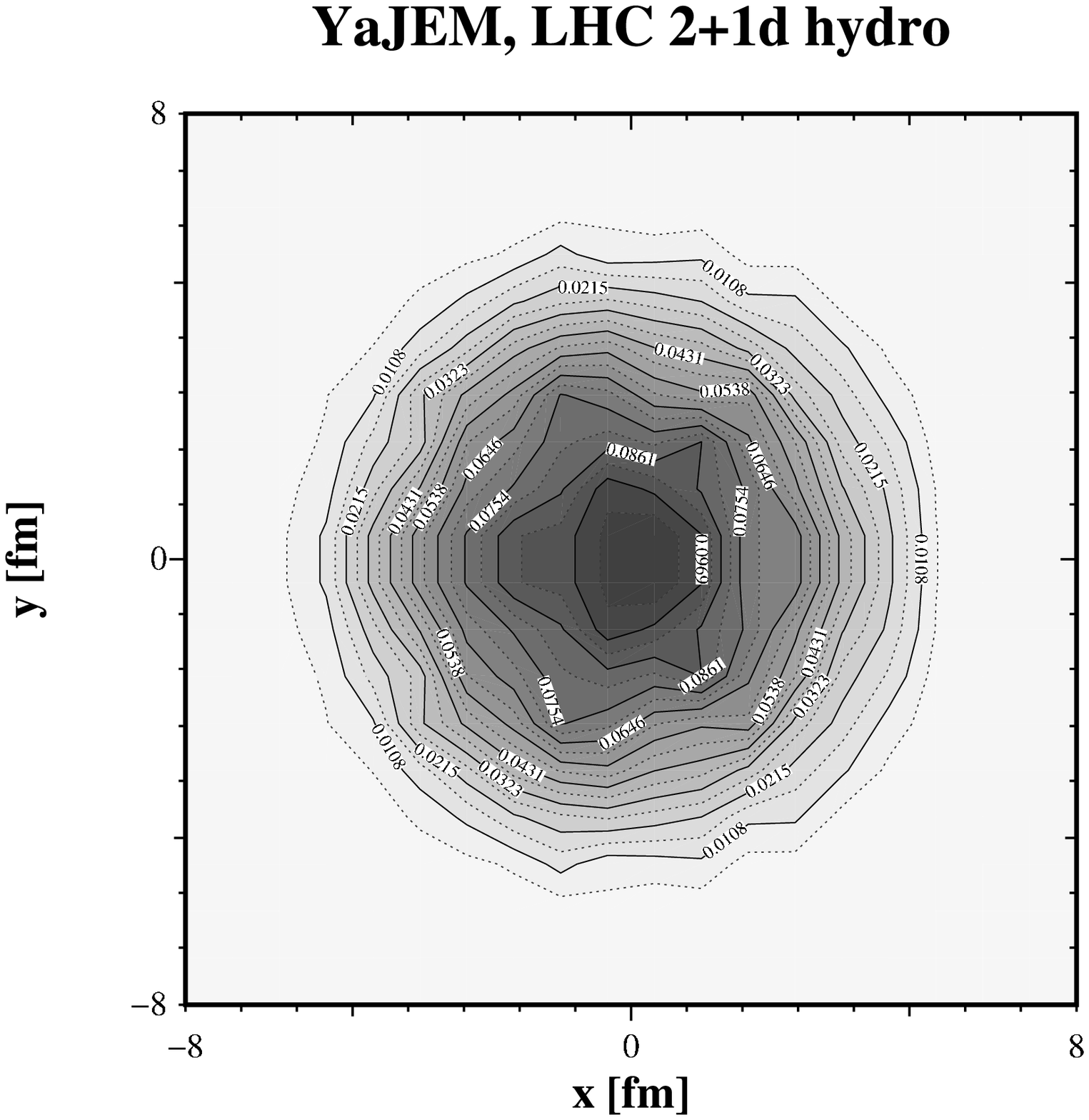, width=8cm}
\end{center}
\caption{\label{F-Vertices}(Color online) The distribution of vertices leading to triggered events (here $-x$ denotes the direction of the trigger jet).}
\end{figure}

The relative robustness against medium modification translates directly into a lack of surface bias. This is illustrated by Fig.~\ref{F-Vertices} which shows the probability distribution of the location of a vertex in the transverse plane given a triggered event.

While the peak of such distributions is usually shifted off-center by $\sim 4$ fm in hadron-triggered correlations, no surface bias is visible in the case of dijet correlations. In other words, while the strength of medium modification and the in-medium pathlength are correlated in dijet as well as dihadron correlations, the strength of the correlation is much weaker due to the relative robustness of jets vs. single hadrons. The presence of strong uncorrelated background fluctuations $O(10-20)$ GeV in the jet area (with the typical scale of energy transported out of cone of the same order \cite{Collimation}) further dilutes the position - energy correlation and effectively removes any surface bias. As a result, the observable is insensitive to pathlength dependence or medium geometry.

\subsection{Generic fate of soft gluons}

While the suppression of single hadron production can be achieved by fairly collinear gluon emission, the suppression of a jet rate requires the emission of large angle gluons to shuffle energy out of the jet cone. Applying this argument to medium-induced gluon production above a thermal momentum scale explains the robustness mentioned above. However, as pointed out in \cite{Collimation}, this argument no longer applies to soft gluons in the thermal momentum range, as they are easily scattered to different angles. This behaviour is fairly generic: whenever a model can produce soft gluons in the shower, these soft gluons are allowed to re-interact with the medium and there is energy and momentum exchange between shower and medium, the soft gluon angles becomes randomized no matter what dynamics is assumed in detail (in YaJEM for instance this effect has been observed the distortion of low $P_T$ jet shapes long before $A_J$ was measured \cite{JetShapes}). In essence, once a gluon in the thermal momentum range is embedded into a thermal medium and interacts with the medium, it is difficult to find any argument why the gluon should not thermalize within a very short time, and so soft gluon production generically equals energy transport out of cone.

We thus end up in a situation in which hard gluon emission is too rare to make an effect (and not governed by shower-medium physics in any case since the momentum scales are too different), the dynamics of semi-soft gluon emission is largely erased by the clustering process into jets and the dynamics of soft gluons is generic and independent of specific model assumptions. This explains why any model which is capable of suppressing the rate of triggered jets expects also a very similar dijet imbalance.

\subsection{The role of background and jet definition}

The above arguments suggest that the overall strength of the medium quenching power is the only jet-medium interaction related parameter constrained by the measured distribution of $A_J$. However, judging from Figs.~\ref{F-Base} and \ref{F-Fluct} the outcome is very sensitive to both the precise cuts used to relate hadrons to calorimeter jets and to the assumed amount of background fluctuations (the latter point has also been made in \cite{BgFluct}). This leads to the surprising conclusion that in an approach which models the whole event, the correct modelling of the initial state fluctuations of the hydrodynamical medium and their evolution is more important than the modelling of shower-medium interactions.

The dependence on the way hadrons are related to calorimeter jets may then explain the somewhat different shapes of the $A_J$ as observed by various groups \cite{Dijets-Qin,Dijets-Vitev,Dijets-Martini}. This should again be taken as a sign of caution not to read too much into the (dis-)agreement of the results presented in this work (and in other works) with the data. But the very observation that the observable is more sensitive to technical details of jet finding and treatment of fluctuations than to the physics it is supposed to measure is clearly a reason to worry.

\section{Conclusions}

Following the reasoning presented in the previous section, it seems that observing the emission of soft gluons at the medium momentum scale is problematic in principle, as such gluons would thermalize under very general conditions. A more meaningful question is then how the energy and momentum carried by such gluons is transported by the bulk medium, i.e. by hydrodynamical excitations.

The physics of medium modifications to a parton shower is then contained in semi-soft gluons with $p_T \sim \text{few } T$ which are sufficiently hard not to be deflected to large angles by re-interactions with the medium. However, the physics at such momentum scales is best not observed via clustering into jets, as the clustering algorithms are designed to suppress sensitivity of observables to physics at these scales. As a consequence, triggerd hadron correlations have much greater tomorgraphic power. 

It would thus appear that fully reconstructed jets do not make the best observables to do high $P_T$ tomography of the medium or to determine the precise nature of shower-medium interaction, and that approches to focus on this physics specifically by suitable designed $n$-particle correlation are more likely to answer such questions.

\begin{acknowledgments}
 
This work was supported by the Academy Researcher program of the Finnish Academy (Project 130472). Discussions with J\"{o}rn Putschke, Helen Caines, Jan Rak, Guangyou Qin, Berndt M\"{u}ller, Jamie Nagle, Brian Cole, Matteo Cacciari and Bj\"{o}rn Schenke are gratefully acknowledged.
 
\end{acknowledgments}

\end{document}